\documentclass[aps,pra,reprint,groupedaddress]{revtex4-1}

\usepackage{amsmath}
\usepackage{graphicx}
\begin{document}

\title{Cobalt in strontium titanate as a new off-center magnetic impurity}


\author{I. A. Sluchinskaya}
\email[]{irinasluch@gmail.com}
\affiliation{Moscow State University, Moscow, 119991 Russia}
\author{A. I. Lebedev}
\affiliation{Moscow State University, Moscow, 119991 Russia}

\date{\today}

\begin{abstract}
The local structure and oxidation state of the cobalt impurity in SrTiO$_3$ is
studied by X-ray absorption fine structure (XAFS) spectroscopy. The synthesis
conditions, under which cobalt predominantly (up to 76\%) substitutes the
atoms at the $A$ site of the perovskite structure, is found for SrTiO$_3$(Co)
samples. By varying the synthesis conditions, it is possible to appreciably
change the ratio between the concentrations of cobalt atoms incorporated into
the $A$ and $B$~sites. It is established that the oxidation state of cobalt is
+2 at the $A$~site and +3 at the $B$~site. It is revealed that the Co impurity
at the $A$~site is off-center, and its displacement from the lattice site is
$\sim$1.0~{\AA}. First-principles calculations show that an isolated Co$^{3+}$ ion
at the $B$~site is diamagnetic, whereas the Co$^{2+}$ ion at the $A$~site
is in a high-spin state ($S = 3/2$).

\medskip
\texttt{Physics of the Solid State 61, 390 (2019); DOI: 10.1134/S1063783419030302}
\end{abstract}


\maketitle

\section{INTRODUCTION}

The search for new magnetic off-center impurities, which can simultaneously result in the
appearance of ferroelectric and magnetic ordering and the magnetoelectric interaction in
crystals~\cite{PhysRevLett.101.165704}, is currently a topical problem. The materials
characterized by these properties are classified as multiferroics---multifunctional materials,
which open new opportunities for modern electronics and spintronics.

It is obvious that to obtain multiferroic properties, a material must have a magnetic moment,
e.g., as a result of its doping with magnetic impurities. The position of an impurity atom
in a crystal and its local environment may have an appreciable effect on its magnetic moment.
In addition, atoms of $3d$ transition elements can exist in several oxidation states, which
depend on the position of an atom in a lattice, its local environment (isolated atom, complexes
with different defects), and the presence of other donors and acceptors in samples.

The oxidation and spin states can be changed by varying the synthesis conditions. Thus, in the
case of the Mn impurity in SrTiO$_3$, it was shown that the oxidation state and the ratio between
the concentrations of atoms at different lattice sites can be changed by varying the annealing
temperature and the stoichiometry of samples (see \cite{JETPLett.89.457,BullRASPhys.74.1235}
and the references therein). Our studies of nickel-doped strontium titanate showed that the
magnetic state of nickel depends on what kind of impurity complexes are formed in the doping
process. The magnetic moment of nickel in the Ni$^{2+}$--$V_{\rm O}$ complex depends on the
impurity--vacancy distance (the nearest or distant vacancy)~\cite{JAdvDielectrics.3.1350031,PhysSolidState.56.449}.

Recent experiments have shown that SrTiO$_3$ doped with cobalt with a concentration of 14--40\%
is a ferromagnet at 300~K and, at the same time, a dielectric~\cite{NewJPhys.12.043044,PhysRevB.87.144422},
which is a rather rare combination. This was one of the reasons for the present study. The preliminary
study of this material suggested a possible incorporation of the Co impurity into different
sites of the perovskite structure~\cite{FundamPhysFerr.2016.211}.

Cobalt-doped strontium titanate has been studied for a rather long time. The similarity of the
crystal structures of SrTiO$_3$ and SrCoO$_{3-x}$ (oxygen-deficient perovskite) suggests the
existence of a continuous series of solid solutions in this system. According to
Ref.~\cite{JPowerSources.210.339}, the solubility of cobalt at the $B$~site of SrTiO$_3$ is
at least 40\%, whereas the authors~\cite{InorgChem.43.8169} reported the preparation of
polycrystalline samples containing up to 90\% of Co using the solid-phase synthesis.
The X-ray studies performed in~\cite{InorgChem.43.8169} revealed the appearance of additional
peaks in X-ray diffraction patterns at $x > 0.5$; these peaks were associated with the
oxygen vacancies ordering. In all earlier published papers, it was presumed that the cobalt
atom substitutes the titanium one. No information about the incorporation of cobalt into the
$A$~site has been reported.

The oxidation state of cobalt in SrTiO$_3$(Co) was determined by electron paramagnetic resonance
(EPR), X-ray photoelectron spectroscopy (XPS), optical absorption, and X-ray absorption near-edge
structure (XANES) spectroscopy. According to the EPR data~\cite{JPhysC.16.5491}, a signal from
Co$^{4+}$ ions in a low-spin state ($S = 1/2$) was observed in single crystals doped with 0.2\%
Co, and the signal intensity strongly increased after preliminary illumination. It seems that
the cobalt impurity in this sample is diamagnetic in the initial state and is in the +3
oxidation state. Indeed, the studies of the optical absorption spectra of single
crystals~\cite{JPhysC.16.5491,CrystReports.49.469} have revealed the absorption lines typical
of the Co$^{3+}$ ion in an octahedral environment. The XPS studies of thin films grown by
molecular-beam epitaxy (MBE) at 550$^\circ$C with a cobalt concentration from 10 to
50\%~\cite{PhysRevB.87.144422} brought to a conclusion that the cobalt oxidation state is 2+.
The conclusion about the same oxidation state was also made from XPS studies of SrTiO$_3$(Co)
nanofibers, which had a cobalt concentration up to 20\% and were prepared by electrospinning
and annealed at 650$^\circ$C~\cite{JMaterSci.47.8216}, and of the samples with 30\% Co grown
by pulsed laser deposition (PLD)~\cite{NewJPhys.12.043044}. The analysis of XPS data enabled
the authors~\cite{NewJPhys.12.043044} to suppose that the impurity Co atoms replace Ti ions
with the formation of oxygen vacancies, and the oxidation state of cobalt in films is lower than
in bulk samples, as the films were grown at a low partial pressure of oxygen. (Since the oxidation
state of cobalt at the $B$~site differs from the titanium ion charge, this implies the
existence of compensating defects in the samples, in particular, oxygen vacancies.)

It should be noted that as the spread of photoelectron peak energies for the Co atom exceeds
their chemical shift upon a change in the oxidation state~\cite{NIST-XPS-DB}, the
conclusion~\cite{NewJPhys.12.043044,PhysRevB.87.144422} about its 2+ oxidation state in
SrTiO$_3$ was made from the appearance of a strong high-energy satellite peak. However, the
same peak with a slightly lower intensity is also present in the samples with trivalent
cobalt~\cite{PhysRevB.46.9976}, so the reliability of this oxidation state determination
method is questionable. An indirect argument for the appearance of Co$^{2+}$ in films
synthesized by PLD may be an increase in the out-of-plane lattice constant with increasing
cobalt concentration~\cite{NewJPhys.12.043044}. This effect obviously differs from the decrease
in the lattice constant observed when doping the bulk crystals~\cite{InorgChem.43.8169,JMagnMagnMater.305.6}.
Unfortunately, no conclusions about the oxidation state of cobalt have been drawn from the XANES
spectra recorded on SrTiO$_3$(Co) films~\cite{NewJPhys.12.043044}.

To explain the origin of the room-temperature ferromagnetism, a number of theoretical
studies~\cite{PhysRevB.87.144422,ApplPhysLett.100.252904,PhysRevB.90.125130}, in which the
nature of this phenomenon was discussed assuming the presence of divalent cobalt in the
samples, have been performed. These studies have demonstrated an essential role of oxygen
vacancies, which form Co$^{2+}$ ion--nearest vacancy complexes with a nonzero magnetic moment.

The studies of the magnetic properties of cobalt-doped SrTiO$_3$ samples have shown that
the bulk synthesis methods do not usually result in the appearance of ferromagnetism at
room temperature~\cite{InorgChem.43.8169,JMagnMagnMater.305.6}. Instead, a quite different
paramagnetic--antiferromagnetic phase transition was observed in single-phase SrTi$_{1-x}$Co$_x$O$_3$ samples with $x = {}$0.35--0.50 at
15--26 K~\cite{JMagnMagnMater.305.6}. The X-ray diffraction studies of these
samples reduced in a 10\% H$_2$--Ar atmosphere revealed
the presence of metallic Co in them. In contrast, thin
cobalt-doped SrTiO$_3$ films grown by the PLD or MBE
methods with dielectric or semiconducting properties
exhibit ferromagnetism at 300~K at a high impurity
concentration ($>$14\%)~\cite{NewJPhys.12.043044,PhysRevB.87.144422,ApplPhysLett.89.012501}.
Cobalt-doped thin
(La,Sr)TiO$_3$ films, which were annealed in a reducing
atmosphere and had metallic conductivity, demonstrate ferromagnetism at 300~K at a much lower cobalt
content ($\sim$2\%)~\cite{ApplPhysLett.83.2199,PhysRevLett.96.027207}.

It should be noted that the properties of samples strongly depend on the method of their
preparation. The samples prepared at high temperatures are characterized by the presence
of trivalent cobalt and the absence of ferromagnetism. On the contrary, divalent cobalt
and ferromagnetism at 300~K appear under conditions, which can be considered as far from
equilibrium (hydrothermal synthesis, PLD, nanofibers). This makes the problem of determining
the real structure of impurity defects formed by cobalt very topical.

In this work, the structural position and oxidation state of magnetic Co impurity in SrTiO$_3$
samples prepared by solid-phase synthesis with different deviations from
stoichiometry are studied by XAFS spectroscopy. It is shown that, depending on the preparation
conditions, the cobalt atoms can predominantly enter either the $B$~sites of the perovskite
structure in the trivalent state, or the $A$~sites in the divalent state. In the latter
case, the impurity is off-center and is in the high-spin state. The obtained results are
compared with the results of the first-principles calculations of the defect structure.

\section{SAMPLES, EXPERIMENTAL AND CALCULATION TECHNIQUES}

Cobalt-doped SrTiO$_3$ samples with an impurity concentration of 2--3\% and different
deviations from stoichiometry were prepared by solid-phase synthesis.
The initial components were SrCO$_3$, nanocrystalline TiO$_2$ synthesized by the hydrolysis
of tetrapropylorthotitanate and dried at 500$^\circ$C, and Co(NO$_3$)$_2$$\cdot$6H$_2$O.
The components were weighed in required proportions, ground in acetone, and annealed in air
in alumina crucibles at 1100$^\circ$C for 4~h. The resulting powders were ground again
and repeatedly annealed under the same conditions for 4~h. Some samples were additionally
annealed in air at 1500 or 1600$^\circ$C for 2~h. To incorporate cobalt into the $A$ or
$B$~site of the perovskite structure, the composition of the samples was deliberately
deviated from stoichiometry towards excess titanium or strontium. All the prepared samples
were dark brown in color.

Extended X-ray absorption fine structure (EXAFS) and X-ray absorption near-edge structure
(XANES) spectra were recorded in fluorescence mode at the Co $K$-edge (7.709~keV) and 300~K
on the KMC-2 station of the BESSY synchrotron radiation source (Germany). Radiation
was monochromatized with a double-crystal Si$_{1-x}$Ge$_x$ (111)-oriented monochromator.
The intensity of radiation incident on a sample was measured with an ionization chamber,
whereas the intensity of the fluorescence radiation was measured with a R{\"O}NTEC
energy-dispersive silicon drift detector.

The EXAFS spectra were analyzed with a widely used IFEFFIT software package~\cite{IFEFFIT}.
The EXAFS function $\chi(k)$ (here, $k = \sqrt{2m(E - E_0)}/\hbar$ is the wave vector of
a photoelectron and $E_0$ is the absorption edge energy) was extracted from the experimental
spectra using the \texttt{ATHENA} software and fitted with the \texttt{ARTEMIS} software
to the theoretical curve calculated for a chosen structural model
    \begin{equation}
    \begin{split}
    \chi(k) &= - \frac{1}{k} \sum_j \frac{N_j S_0^2}{R_j^2} |f_j(k)| \\
    &\times \exp[-2 \sigma_j^2 k^2 - 2R_j/\lambda(k)] \sin [2kR_j + \phi_j(k)], \\
    \end{split}
    \end{equation}
where the sum runs over several nearest shells $j$ about the central Co atom, $N_j$ is the
number of atoms in the $j$th coordination shell, $R_j$ is its radius, $f_j$ is the scattering
amplitude, $\lambda$ is an effective mean free path, $\sigma_j^2$ is the Debye--Waller
factor characterizing the mean-squared deviation between the distance to atoms of the
$j$th sort from its mean value, and $S_0^2$ is the multiplier taking into account
the reduction in the oscillation amplitude resulting from many-body effects. The scattering
amplitude $f_j(k)$, the phase shifts $\phi_j(k)$, and the mean free path $\lambda(k)$ for
all single- and multiple-scattering paths were calculated using the FEFF6 software.
For each sample, 3--4 spectra were recorded and further independently processed, and the
obtained curves $\chi(k)$ were averaged. The details of the data analysis are given
in~\cite{PhysRevB.55.14770}.

The first-principles calculations of the geometry and magnetic and electronic structures of
impurity centers were performed using the \texttt{ABINIT} software within the LDA+$U$ approach
on 80-atom cubic supercells, in which one of the Ti$^{4+}$ ions at the $B$~site or one of
the Sr$^{2+}$ ions at the $A$~site of the perovskite structure was
substituted by the cobalt ion. The atoms with partially filled $d$ shell were described using
the projector-augmented wave (PAW) pseudopotentials. The parameters describing the Coulomb
and exchange interactions inside the $d$ shell were $U = 5$ and $J = 0.9$~eV.

\section{EXPERIMENTAL RESULTS}
\label{sec3}

\begin{figure}
\includegraphics{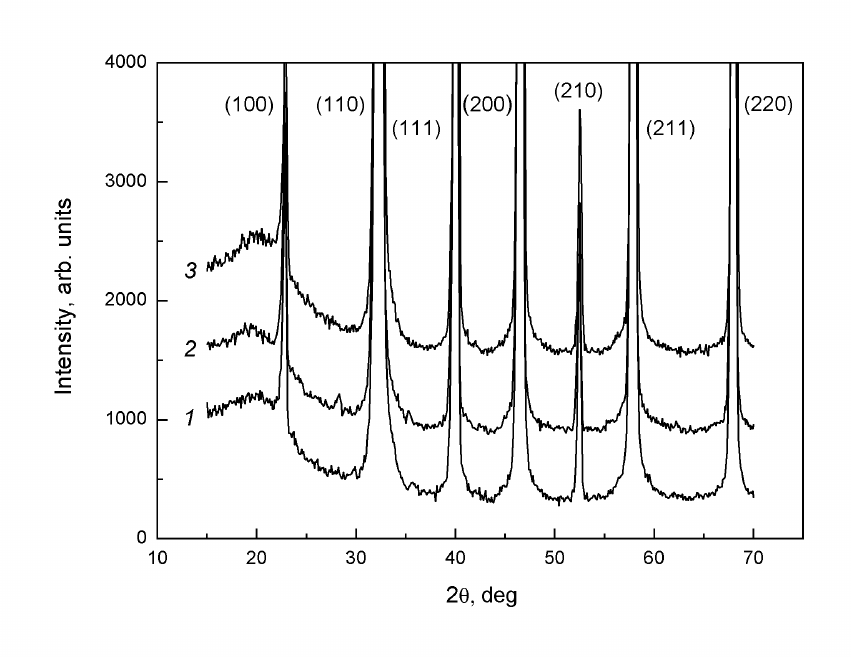}
\caption{\label{fig1} X-ray diffraction patterns for the samples (1)
Sr$_{0.98}$Co$_{0.02}$TiO$_3$ annealed at 1600$^\circ$C, (2) Sr$_{0.98}$Co$_{0.02}$TiO$_3$
annealed at 1500$^\circ$C, (3) SrTi$_{0.97}$Co$_{0.03}$O$_3$ annealed at
1500$^\circ$C.}
\end{figure}

The X-ray diffraction analysis of prepared samples showed that all studied SrTiO$_3$(Co)
samples were single-phase and had a cubic perovskite structure at 300~K. Typical X-ray
diffraction patterns of three samples are shown in Fig.~\ref{fig1}, and their unit cell
parameters together with the unit cell parameter of undoped SrTiO$_3$ are given in
Table~\ref{table1}. A decrease in the unit cell parameter of all doped specimens in
comparison with undoped SrTiO$_3$ agrees with the earlier published
data~\cite{InorgChem.43.8169,JMagnMagnMater.305.6} and indicates the formation of a solid
solution.

\begin{figure}
\includegraphics{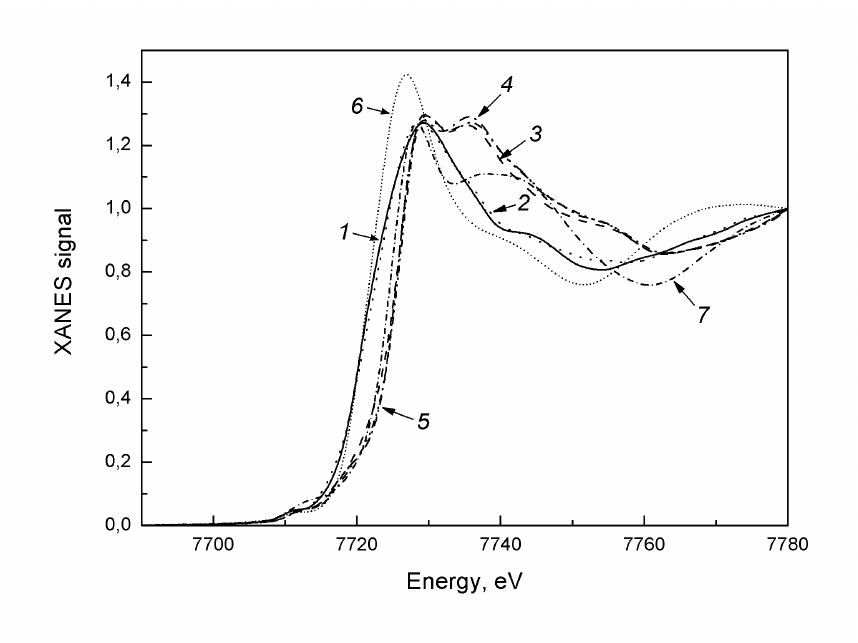}
\caption{\label{fig2} XANES spectra for SrTiO$_3$(Co) samples and reference cobalt compounds:
(1) Sr$_{0.98}$Co$_{0.02}$TiO$_3$ annealed at 1600$^\circ$C, (2) Sr$_{0.98}$Co$_{0.02}$TiO$_3$
annealed at 1500$^\circ$C, (3) Sr$_{0.98}$Co$_{0.02}$TiO$_3$ annealed at 1100$^\circ$C, (4)
SrTi$_{0.97}$Co$_{0.03}$O$_3$ annealed at 1500$^\circ$C, (5) SrTi$_{0.97}$Co$_{0.03}$O$_3$
annealed at 1100$^\circ$C, (6) Co(NO$_3$)$_2$$\cdot$6H$_2$O, (7) LaCoO$_3$.}
\end{figure}

The oxidation state of the Co impurity in doped SrTiO$_3$ samples was determined by comparing
the absorption edge position in the XANES spectra of the samples with the edges positions
in the reference samples LaCoO$_3$ (cobalt valence +3) and Co(NO$_3$)$_2$$\cdot$6H$_2$O
(cobalt valence +2). The XANES spectra of all samples and reference compounds are shown in
Fig.~\ref{fig2}.

The comparison of the XANES spectra of all SrTiO$_3$(Co) samples annealed at a temperature
of 1100$^\circ$C and the sample annealed at 1500$^\circ$C with deviation from stoichiometry
towards excess strontium showed that the absorption edges in these samples are close to
each other and almost coincide with the absorption edge of the LaCoO$_3$ reference compound.
This suggests that the Co impurity in these samples is predominantly in the +3 oxidation
state.

\begin{table*}
\caption{\label{table1} Unit cell parameters for three SrTiO$_3$(Co) samples and undoped SrTiO$_3$}
\begin{tabular}{ccccc}
\hline
                               & Sr$_{0.98}$Co$_{0.02}$TiO$_3$ & Sr$_{0.98}$Co$_{0.02}$TiO$_3$ & SrTi$_{0.97}$Co$_{0.03}$O$_3$ & \\
\smash{\raisebox{7pt}{Sample}} & annealed at 1600$^\circ$C     & annealed at 1500$^\circ$C     & annealed at 1500$^\circ$C     & \smash{\raisebox{7pt}{SrTiO$_3$}} \\
\hline
Unit cell parameter $a$~({\AA}) & 3.8977$\pm$0.0005 & 3.8994$\pm$0.0002 & 3.8931$\pm$0.0006 & 3.905 \\
\hline
\end{tabular}
\end{table*}

The absorption edges of SrTiO$_3$(Co) samples annealed at 1500 and 1600$^\circ$C with deviation
from stoichiometry towards excess Ti coincide with each other and are close to the absorption
edge of the Co(NO$_3$)$_2$$\cdot$6H$_2$O reference compound. This indicates that these samples
contain cobalt predominantly in the +2 oxidation state.

It should be noted that the chemical shift in the XANES spectra is much larger than that in the
photoelectron spectra~\cite{NIST-XPS-DB}, thus providing the possibility to make a more reliable
conclusion about the oxidation state of cobalt.

To determine the structural position of the Co impurity in SrTiO$_3$, the EXAFS spectra were
analyzed. The data analysis and the choice of a structural model were performed taking into
account the results of X-ray diffraction and XANES studies, namely, the unit cell parameter
and the ability of Co atoms to enter different ($A$ and $B$) sites of the perovskite structure,
probably simultaneously and in different oxidation states.

The comparison of EXAFS spectra for two groups of samples, in which Co ions are in different
oxidation states, reveals a qualitative distinction between them.

For all samples in which Co is in the Co$^{3+}$ oxidation state, a reasonable agreement between
the experimental spectra and the theoretically calculated curves $\chi(k)$ was obtained for
the model, in which the Co$^{3+}$ ion substitutes the Ti$^{4+}$~ion with the formation of a distant
oxygen vacancy. In this model, the Co--O distance in the first shell was $1.909 \pm 0.012$~{\AA},
the Co--Sr distance was $3.342 \pm 0.016$~{\AA}, and the Co--Ti distance was $3.892 \pm 0.012$~{\AA}.

The analysis of EXAFS spectra for the Sr$_{0.98}$Co$_{0.02}$TiO$_3$ samples annealed at 1600
and 1500$^\circ$C, in which the impurity Co atom is in the +2 oxidation state, has shown that
none of the models, in which the Co$^{2+}$~ion substitutes either Ti$^{4+}$ at the $B$~site
or Sr$^{2+}$ at the $A$~site, agree satisfactorily with the experiment. Qualitative agreement
between the experimental spectrum and the theoretically calculated EXAFS curve for these samples
was obtained for the model, in which the Co$^{2+}$ ion replaces the Sr$^{2+}$ ion at the $A$~site
and is displaced from this site in the [100] direction by $\sim$1.0~{\AA}. This displacement seems
to be reasonable taking into account a large difference between the Co$^{2+}$ and Sr$^{2+}$
ionic radii. In this case, the distance from the cobalt atom to the nearest oxygen atoms is
$1.993 \pm 0.042$~{\AA}.

Unfortunately, the agreement between the experimental and theoretically calculated EXAFS curves
was not sufficiently good for all studied samples (the criterion of agreement between the
calculated and experimental spectra was a low $R$-factor, the correspondence of obtained
coordination numbers to the theoretical model, and the agreement between the Fourier transforms
of spectra in the $R$-representation). For this reason, it was decided to consider the microscopic
models of different impurity centers with the Co atom at the $A$ and $B$~sites.

\section{RESULTS OF FIRST-PRINCIPLES CALCULATIONS}
\label{sec4}

The construction of theoretical models was performed taking into account the data on the oxidation
state of impurity atoms and the interatomic distances obtained from the analysis of EXAFS spectra.
We note that the earlier first-principles calculations~\cite{PhysRevB.87.144422,ApplPhysLett.100.252904,PhysRevB.90.125130}
were performed mainly for the divalent cobalt at the $B$~site, while our results indicate other
oxidation states and lattice sites.

The modeling of the geometry of impurity centers was started from the Co$^{2+}$ ion at the
$A$~site. This impurity center is one of the simplest ones, as the substitution of strontium
by cobalt does not require any charge compensation. The calculations, which took into account
the full relaxation of the local environment, showed that the on-center position of the
Co$^{2+}$ ion at the $A$~site in SrTiO$_3$ is energetically unstable, and the impurity atom
displaces to an off-center position. The comparison of the energies of structures with atoms
displaced along the [100], [110], and [111] directions showed that the structure with the
displacement along the [110] axis has the lowest energy, and the energy of the structures
with the displacements along the [100] and [111] axes are, respectively, by 264 and 985~meV
(per Co atom) higher. The displacements along the [100], [110], and [111] axes are 1.03,
1.00, and 0.79~{\AA}, respectively, and the ion magnetic moment is $S = 3/2$ in all
configurations. The calculated distances to the nearest atoms in the cobalt environment for
these configurations are given in Table~\ref{table2}. It is seen that the calculated distances
to the nearest oxygen atoms appreciably exceed the Co--O distances for cobalt at the $B$~site,
and the interatomic distances for the displacements along the [100] and [111] axes are closest
to the experimentally obtained value of $1.993 \pm 0.042$~{\AA}.

\begin{table*}
\caption{\label{table2} Local environment of Co ions in different theoretical models. The distances are in~{\AA}.}
\begin{tabular}{cccccccc}
\hline
 &  & \multicolumn{6}{c}{Shell} \\
\cline{3-8}
\smash{\raisebox{7pt}{Model}}    & \smash{\raisebox{7pt}{$S$}} & 1 & 2 & 3 & 4 & 5 & 6 \\
\hline
Off-center atom at the $A$ site, & 3/2 & 2.014 & 2.948 & 3.056 & 3.168 & 3.630 & 3.814 \\
displacement [100]               &     & (4O)  & (4Ti) & (1Sr) & (4O)  & (4O)  & (4O) \\

Off-center atom at the $A$ site, & 3/2 & 2.051 & 2.069 & 2.706 & 3.265 & 3.318 & 3.474 \\
displacement [110]               &     & (1O)  & (4O)  & (2Ti) & (2O)  & (2Sr) & (4Ti) \\

Off-center atom at the $A$ site, & 3/2 & 1.996 & 2.707 & 2.946 & 3.216 & 3.523 & 3.564 \\
displacement [111]               &     & (3O)  & (1Ti) & (6O)  & (3Ti) & (3Sr) & (3O) \\

Isolated Co$^{3+}$ ion at the $B$ site & 0 & 1.905 & 3.352 & 3.801 & 4.355 & 5.509  & \\
                                       &   & (6O)  & (8Sr) & (6Ti) & (24O) & (12Ti) & \\
\hline
\end{tabular}
\end{table*}

The modeling of impurity centers with trivalent cobalt at the $B$~site implies the existence
of defects compensating the difference between the ionic charges. The calculations for an
isolated Co$^{3+}$~ion at the $B$~site, for which an additional electron was provided by a
distant F donor atom located at the oxygen site (Table~\ref{table2}), have shown that the
low-spin state ($S = 0$) has the lowest energy. The comparison of the distances obtained
in the model, in which the Co$^{3+}$ ion enters the $B$~site, with the experimentally
determined distance of $1.909 \pm 0.012$~{\AA} shows their good agreement.

\begin{figure}
\includegraphics{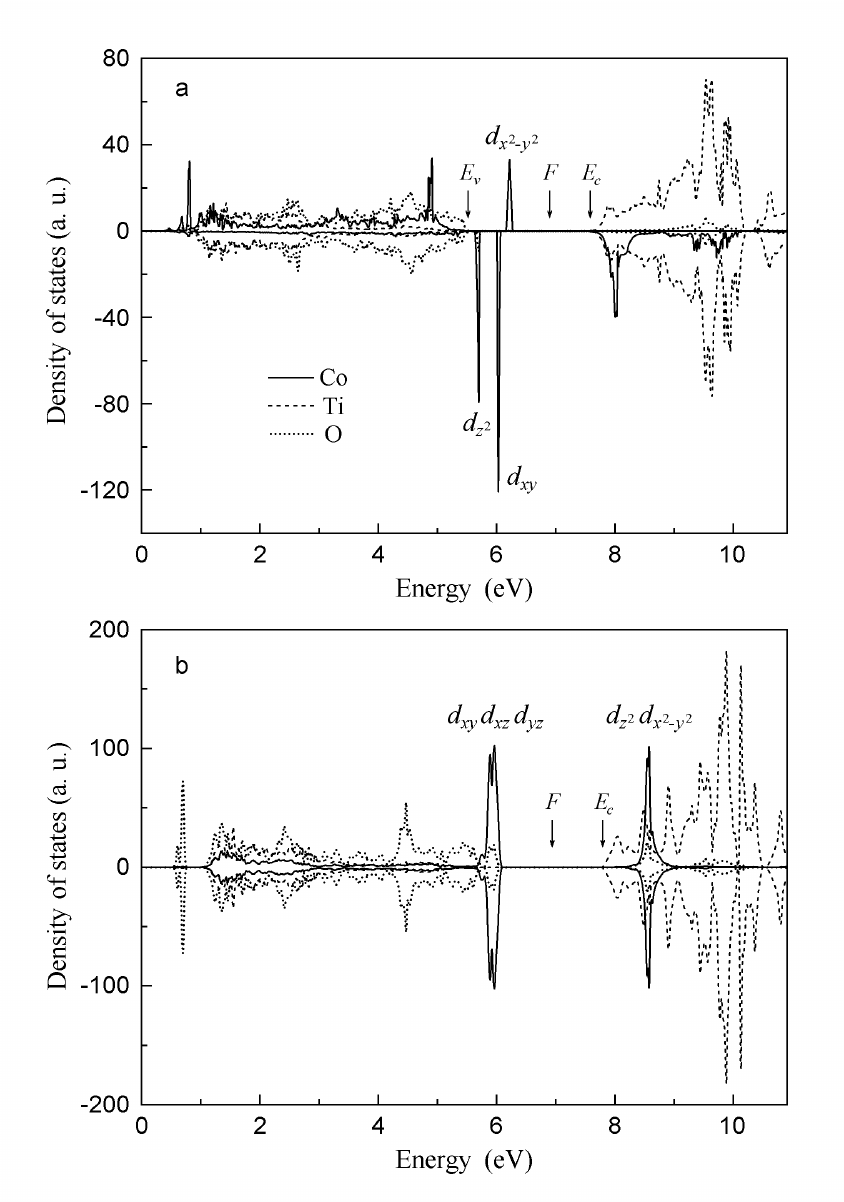}
\caption{\label{fig3} Partial contributions of cobalt, titanium, and oxygen to the density of
states for (a) Co$_A^{2+}$ and (b) Co$_B^{3+}$ impurity centers. Partial contribution of cobalt
was decreased by 5~times.}
\end{figure}

The electronic structure of the Co$^{2+}$ impurity center at the $A$~site is shown in Fig.~\ref{fig3}(a). It is seen that the
cobalt impurity creates three acceptor levels associated
with the occupied $d_{x^2-y^2}$ (spin up) and $d_{z^2}$ and $d_{xy}$ (spin
down) orbitals in the forbidden gap of SrTiO$_3$. The
Fermi level $F$ is located in the forbidden gap. The $d_{z^2}$
state is strongly hybridized with the oxygen $p$ states, so
the optical transitions allowed by the selection rules
from this level to the bottom $E_c$ of the conduction band
(which is formed by cobalt and titanium $d$ states) may
be the cause of intense color of the samples.

For an isolated Co$^{3+}$ impurity center at the $B$~site, the cobalt $d$ levels of
$T_{2g}$ symmetry merge together with the valence band edge, whereas the levels of $E_g$
symmetry form resonant levels in the conduction band (Fig.~\ref{fig3}(b)). That is why there
are no reasons to expect appreciable change in the color of samples in this case.

\section{REFINEMENT OF THE STRUCTURAL MODELS AND DISCUSSION OF RESULTS}

Insufficiently good agreement obtained between the experimental and calculated EXAFS spectra
in Sec.~\ref{sec3} has stimulated us to consider more complicated models. First, we considered
the model, in which cobalt at the $B$~site forms a complex with the nearest oxygen vacancy.
The distances obtained in this model for the SrTi$_{0.97}$Co$_{0.03}$O$_3$ sample annealed at
1100$^\circ$C are only slightly different from the values given above, but the parameter
$S_0^2$ was closer to its value for the reference compounds.

Using the geometry of the local environment of the off-center cobalt atom at the $A$~site
calculated for three directions of displacement, the spectrum of the Sr$_{0.98}$Co$_{0.02}$TiO$_3$
sample annealed at 1600$^\circ$C was processed. It turned out that the lowest $R$ factor is
obtained by fitting curves for configurations with the [100] and [111] displacements. The
calculated Fourier transform of the EXAFS spectrum for the model with the [110] displacement
had a strong peak from Ti atoms in the third shell, but this peak was entirely absent in the
experimental spectrum. For this reason, the [110] displacement may be completely excluded
from further consideration. Taking into account an appreciable difference in energy between
the configurations with the [100] and [111] displacements, we believe that the displacement
of atoms occurs in the [100] direction and shall consider only this model in what follows.

Further, we have supposed that cobalt atoms in actual samples may substitute atoms at
the $A$ and $B$~sites simultaneously. The structural parameters for two EXAFS spectrum
components, which will be called states $A$ and $B$ and in which the impurity atoms substitute
the corresponding lattice sites, were determined as follows. As a starting point, it was assumed
that the spectrum of the SrTi$_{0.97}$Co$_{0.03}$O$_3$ sample annealed at 1100$^\circ$C
represents the spectrum of ``pure'' state~$B$. The local environment of cobalt at the $B$~site
was modeled taking into account the nearest oxygen vacancy. For this model, a set of structural
parameters (distances and Debye--Waller factors for three nearest shells) was calculated.
The spectrum of the Sr$_{0.98}$Co$_{0.02}$TiO$_3$ sample annealed at 1600$^\circ$C was then
presented as a sum of states $A$ and $B$ with unknown relative fractions. Fixing the parameters
determined for state $B$ at the previous step and taking the interatomic distances predicted
by first-principles calculations for state $A$ (see Sec.~\ref{sec4}) as another starting point,
we estimated the structural parameters for state $A$ and the relative fractions of states
$A$ and $B$. Further, assuming that the SrTi$_{0.97}$Co$_{0.03}$O$_3$ sample annealed at
1100$^\circ$C may contain a small fraction of state $A$, we performed a new analysis of the
EXAFS spectrum of this sample. At this step, the data obtained for state $A$ at the previous
step were fixed. This gave the refined parameters for state $B$ and the relative fractions of
states $A$ and $B$ in this sample. The multiple repetition of the described iterative
procedure provided the structural parameters for states $A$ and $B$ and their relative fractions
in the analyzed spectra.

\begin{figure}
\includegraphics{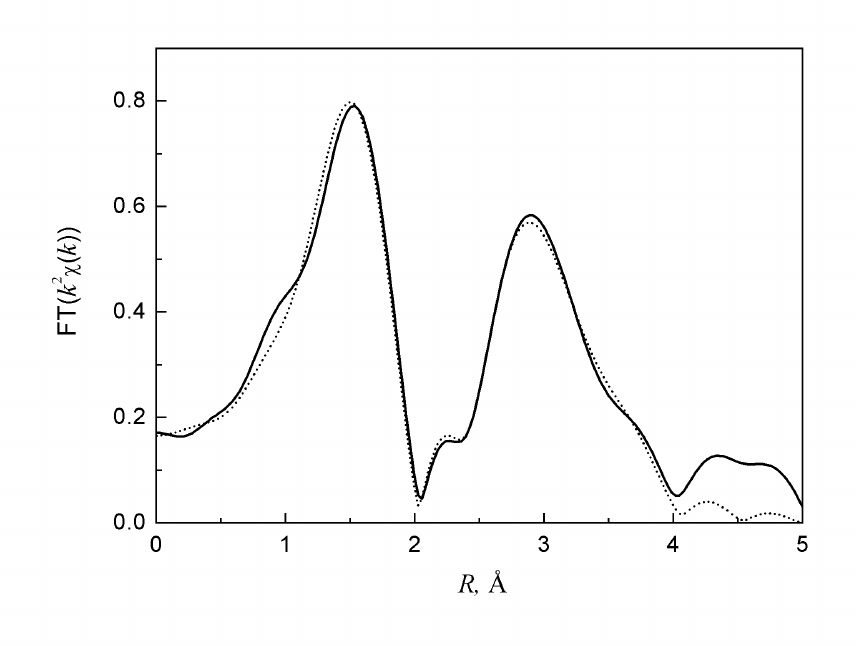}
\caption{\label{fig4} Comparison of Fourier transform amplitudes for a typical EXAFS spectrum
of the Sr$_{0.98}$Co$_{0.02}$TiO$_3$ sample annealed at 1600$^\circ$C (solid line) with
the calculated curve (dotted line) corresponding to the best fit.}
\end{figure}

A typical experimental EXAFS spectrum and its best theoretical fit within the described
approach are shown in Fig.~\ref{fig4}, and the interatomic distances to the nearest shells
for samples annealed at 1100 and 1600$^\circ$C are given in Table~\ref{table3}. The
interatomic distances obtained from experiment agree with their calculated values (compare
Tables~\ref{table2} and \ref{table3}). According to the iterative procedure results, the
Sr$_{0.98}$Co$_{0.02}$TiO$_3$ sample annealed at 1600$^\circ$C contains 76\% of the
incorporated cobalt at the $A$~site, whereas the SrTi$_{0.97}$Co$_{0.03}$O$_3$ sample
annealed at 1100$^\circ$C contains nearly 18\% of impurity atoms at the $A$~site.

\begin{table*}
\caption{\label{table3} Structural parameters determined from the EXAFS spectra analysis of two
studied samples}
\begin{tabular}{ccccccc}
\hline
Sample & Model & $R$-factor & $S_0^2$ & Shell & $R_i$~({\AA}) & $\sigma_i^2$~({\AA}$^2$) \\
\hline
Sr$_{0.98}$Co$_{0.02}$TiO$_3$  & Site $A$ & 0.00297 & 0.705 & $R_{\rm Co-O}$  & 2.040(6)  & 0.0049(18) \\
annealed at 1600$^\circ$C      &          &         &       & $R_{\rm Co-Ti}$ & 3.094(15) & 0.0146(27) \\
                               &          &         &       & $R_{\rm Co-Sr}$ & 2.816(20) & 0.0114(40) \\
                               & Site $B$ &         & 0.282 & $R_{\rm Co-O}$  & 1.906(11) & 0.0037(18) \\
                               &          &         &       & $R_{\rm Co-Sr}$ & 3.354(19) & 0.0087(12) \\
                               &          &         &       & $R_{\rm Co-Ti}$ & 3.907(17) & 0.0064(13) \\
SrTi$_{0.97}$Co$_{0.03}$O$_3$  & Site $A$ & 0.00255 & 0.153 & $R_{\rm Co-O}$  & 2.040(6)  & 0.0049(18) \\
annealed at 1100$^\circ$C      &          &         &       & $R_{\rm Co-Ti}$ & 3.094(15) & 0.0146(27) \\
                               &          &         &       & $R_{\rm Co-Sr}$ & 2.816(20) & 0.0114(40) \\
                               & Site $B$ &         & 0.731 & $R_{\rm Co-O}$  & 1.906(11) & 0.0037(18) \\
                               &          &         &       & $R_{\rm Co-Sr}$ & 3.354(19) & 0.0087(12) \\
                               &          &         &       & $R_{\rm Co-Ti}$ & 3.907(17) & 0.0064(13) \\
\hline
\end{tabular}
\end{table*}

\mbox{ }

\section{CONCLUSIONS}

In this paper, the results of XAFS spectroscopy study of the local structure and oxidation state
of the cobalt impurity in SrTiO$_3$ have been considered. It was revealed that under certain
synthesis conditions cobalt may be incorporated into the $A$~site of the perovskite structure.
By varying the synthesis conditions, it is possible to appreciably change the ratio of
concentrations of cobalt atoms incorporated into the $A$ and $B$~sites. Thus, 76\% of Co atoms
enter the $A$~site at an annealing temperature of 1600$^\circ$C and only 18\% of them enter
this site at 1100$^\circ$C.

The study of XANES spectra showed that the oxidation state of cobalt is +2 at the $A$~site and
+3 at the $B$~site. It was established that the absorption edge shift in the XANES spectra is
much more sensitive to the change in the oxidation state of cobalt in comparison with XPS.
EXAFS studies revealed that Co atoms at the $A$~site are off-center, and their displacement
is $\sim$1.0~{\AA}. According to first-principles calculations, the magnetic moment of the
Co$^{2+}_A$ ion is $S = 3/2$. Thus, Co$^{2+}_A$ ions in SrTiO$_3$ are another representative
of the group of magnetic off-center impurities, with which interesting multiferroic properties
may be associated.

\begin{acknowledgments}
This work was supported by the Russian Foundation for Basic Research (grant no. 17-02-01068).
\end{acknowledgments}


\providecommand{\BIBYu}{Yu}

\end{document}